\documentclass[11pt,a4paper,oneside]{article}

\usepackage{mymacro-jhep}

\newcommand{\jt}{\tilde{\jmath}}
\newcommand{\Deltat}{\tilde{\Delta}}

\author[a,b]{Andrea Manenti\lsp}

\affiliation[a]{Institute of Physics, \'Ecole Polytechnique F\'ed\'erale de Lausanne, CH-1015 Lausanne, Switzerland}
\affiliation[b]{Simons Center for Geometry and Physics, Stony Brook, NY 11794, USA}

\title{Thermal CFTs in momentum space}

\abstract{We study some aspects of conformal field theories at finite temperature in momentum space. We provide a formula for the Fourier transform of a thermal conformal block and study its analytic properties. In particular we show that the Fourier transform vanishes when the conformal dimension and spin are those of a ``double twist'' operator $\Delta = 2\Delta_\phi + \ell + 2n$. By analytically continuing to Lorentzian signature we show that the spectral density at high spatial momenta has support on the spectrum condition $|\omega| > |k|$. This leads to a series of sum rules. Finally, we explicitly match the thermal block expansion with the momentum space Green's function at finite temperature in several examples.}

\begin{document}

\maketitle

\section{Introduction}
The operator product expansion (OPE) is a remarkably powerful tool of central importance in conformal field theories (CFTs). The convergence of the OPE in flat space \cite{Mack:1976pa,Pappadopulo:2012jk} intuitively follows from the fact that one can always find a sphere enclosing two operators and perform a scale transformation to bring them together. However the story is not so straightforward in other geometries. The most physically interesting case is perhaps $\BBR^{d-1}\times S^1$, representing a CFT that describes a quantum critical point in $d$ space-time dimensions at a temperature proportional to the inverse radius of the circle, or the holographic dual of a black hole in $d+1$ space-time dimensions.
In this thermal geometry---which is the main object of study of this paper---the OPE stops converging when two operators are at a separation larger than the circumference of the $S^1$. This means that the OPE does not give us access to the physics at large distances (or equivalently, at low momentum).
\par
Nevertheless there are many examples where, surprisingly, the OPE provides a good approximation of the correct answer \cite{
Katz:2014rla,Witczak-Krempa:2015pia
}. Establishing rigorously the limits of validity of this approximation in the regime of low frequencies is still an open problem. But we can try to make some progress by understanding better the easier regime of high frequency. The limit $\omega \gg T$, for example, was addressed in \cite{CaronHuot:2009ns} and it was conjectured that it would always be possible to express the Euclidean Green's function as an asymptotic expansion in $1/\omega$.
\par
Here however we consider a different limit, namely of large spatial momentum $|\bfk|\gg T$ and arbitrary $\omega$. Building on the formalism developed in \cite{Iliesiu:2018fao} we are able to express the momentum space Euclidean Green's function of two scalar fields $\phi$ as a sum over the operators in the OPE $\phi\times\phi$. Each operator contributes with a so-called ``thermal conformal block,'' in analogy with the conformal blocks for flat space four-point functions. The advantage of this representation is that it allows us to easily analytically continue to Lorentzian signature $-i\omega_n \to \omega$, where the most interesting physical phenomena take place.
\par
A particularly interesting quantity in Lorentzian signature is the spectral density $\rho(k,\omega)$ defined as the Fourier transform of the commutator $\int \di x\,\langle [\phi(t,\bfx),\phi(0)] \rangle_\beta\, e^{i\bfk\cdot \bfx-i\omega t}$, $\langle\,\cdot\,\rangle_\beta$ representing the expectation value in the thermal ensemble ($\beta = 1/T$). It also has a representation as a sum over the states of the theory.
\eqn{
\rho(k,\omega) = \frac{(2\pi)^d}{Z}\sum_{n,m}\delta(E_n - E_m - \omega) \delta^{d-1}(\mathbf{P}_n-\mathbf{P}_m-\bfk)\big(e^{-\beta E_m}-e^{-\beta E_n}\big) \big|\vev{n|\phi(0)|m}\big|^2
}[eq:rhoasspectralsum]
The states $|n\rangle$ of the CFT are forced to satisfy the condition $E_n \geq |\mathbf{P}_n|$, but this does not necessarily force $\rho(k,\omega)$ to be zero if $|k| > |\omega|$. An intuitive explanation \cite{Banerjee:2019kjh} is that in the thermal background $\phi$ can create a particle with some energy $E$ and momentum $k$ and destroy another one with a similar energy but opposite momentum. This excitation will have a small energy but possibly large momentum. If we compute the spectral density block by block employing the analytic continuation explained before we find that every single term satisfies the spectrum condition (i.e. it is zero for $|k|>|\omega|$). We conclude from this that the spectral density of any CFT will satisfy such condition asymptotically when $k \gg T$, which is the regime where we can trust the OPE. This is to our knowledge a new result. The physical interpretation is clear if one thinks of $k \gg T$ as the low temperature limit: in this case thermal correlators are approximately vacuum expectation values. It can also be seen explicitly by noticing that the terms with $n m \neq 0$ in \eqref{eq:rhoasspectralsum} are subleading when $\beta = 1/T \to \infty$.
\par 
We would also like to emphasize a rather remarkable property of the momentum space conformal blocks. They identically vanish for operators with the so-called ``double twist'' conformal dimension $\Delta = 2\Delta_\phi + \ell + 2n$, where $\Delta_\phi$ is the conformal dimension of the operator whose correlator is being considered. At first one might think that this is an accident due to the fact that the Fourier integrals are not defined for some values of $\Delta$, and ultimately it would lead to a mismatch for theories with double twist operators in the spectrum. It is however perfectly consistent as we observed in various examples presented at the end of the paper.
It turns out that these zeros in the thermal blocks are related to the zeros in the finite temperature inversion formula. This formula was introduced in \cite{Iliesiu:2018fao} in analogy with the Lorentzian inversion formula~\cite{Caron-Huot:2017vep,Simmons-Duffin:2017nub} for the large spin perturbation theory (or lightcone bootstrap)\cite{Fitzpatrick:2012yx,Komargodski:2012ek}. See also \cite{Iliesiu:2018zlz} for a recent application of their method. The search for such a connection was motivated by an apparent puzzle, namely the fact that theories that differ only by the double twist spectrum would look completely identical in momentum space. We will give a simple explanation of its resolution which follows from Kubo-Martin-Schwinger crossing symmetry~\cite{Kubo:1957mj,PhysRev.115.1342}.\footnote{Kubo-Martin-Schwinger crossing is the symmetry of Euclidean thermal correlators under $\tau \to \beta-\tau$, where $\tau$ is the Euclidean time. The inversion formula applied to a block computes its contribution to the crossed channel.\label{footnote:KMS}}
\par
The vanishing of double twist operators has important implications for many theories that admit a perturbative expansion. Large $N$ or the $\epsilon$ expansion are some examples. The reason is that many operators will be nearly double twist at the lowest orders in the expansion parameter, making such theories more efficiently tractable. This is also a strong hint that theories with higher-spin symmetry are generally easier to study since all the conserved currents will disappear when going to momentum space.
\par
It should be mentioned that these zeros also have a drawback. As was observed in \cite{Iliesiu:2018fao} and later generalized in \cite{Petkou:2018ynm}, the thermal mass $m_\mathrm{th}$ of $O(N)$ theories in odd dimensions can be determined by imposing the absence of a certain class of double twist operators from the spectrum. Clearly, since we do not have access to these operators, such a constraint cannot be imposed in momentum space.
\par
The paper is organized as follows: in \sectionname~\ref{sec:FT} and \ref{sec:analyticprop} we compute the Fourier transform of the thermal block and discuss its analytic properties. In \sectionname~\ref{sec:doubletwistinversion} we provide an interpretation of the vanishing of the Fourier transform at the double twist dimensions. In \sectionname~\ref{sec:spectral} we discuss the spectral density (i.e. the Fourier transform of the commutator) and show that at $k\gg T$ it has support over the spectrum condition $|\omega|>|k|$. Then we also prove a series of sum rules inspired to the deep inelastic scattering studied in~\cite{Komargodski:2012ek,Komargodski:2016gci}. \sectionname~\ref{sec:examples} and \ref{sec:ex2d} contain some simple examples that explicitly show how the sum over thermal conformal blocks reproduces the correct answer in momentum space. Finally we leave a few details to the appendices. 

\section{Fourier transform of the thermal blocks}\label{sec:FT}

\subsection{Integral in general dimensions}

Consider the Euclidean two-point function of a scalar field $\phi$ on the geometry $\BBR^{d-1} \times S^1$, where $\tau \sim \tau + 1$ is the compactified time direction. This describes a thermal correlator at inverse temperature $\beta = 1$.\footnote{
The temperature can be set to $1$ without loss of generality. It suffices to do a conformal rescaling of the cylinder $\BBR^{d-1}\times S^1$, which is allowed because it is a flat geometry and thus there are not trace anomalies. The reader can however restore the $\beta$ dependence in all expression by dimensional analysis.}${}^{,}$\footnote{Unlike the case $S^{d-1}\times S^1$ there is no phase transition in $T$. From the holographic point of view we are always in the AdS black hole (or high temperature) phase. We thank Guido Festuccia for this comment.} As discussed in \cite{Iliesiu:2018fao}, when the distance of the two fields lies inside the disk $|x| = \sqrt{\tau^2+\bfx^2} \leq 1$, the correlator can be expressed as an operator product expansion (OPE).
\eqn{
g(\bfx,\tau)\equiv\langle \phi(\bfx,\tau) \phi(\mathbf{0},0)\rangle_\beta = \sum_{\CO \in \phi\times\phi} a_\CO \,C_J^{(\nu)}(\xi)\,|x|^{\Delta-2\Delta_\phi}\,,\qquad \nu = \frac{d-2}{2}\,,\;\;\xi = \frac{\tau}{|x|}\,,
}[eq:euclidOPE]
where $\Delta_\phi$ is the conformal dimension of $\phi$, $C_J^{(\nu)}$ is a Gegenbauer polynomial and the sum is over all primary operators $\CO$ in the $\phi\times\phi$ OPE, whose conformal dimension and spin have been denoted $\Delta$ and $J$.
The function $C^{(\nu)}(\xi) |x|^{\Delta-2\Delta_\phi}$ will be referred to as a \emph{thermal conformal block}. The coefficients $a_\CO$ can be computed in terms of other data of the theory, namely thermal one-point functions $b_\CO$ and two- and three-point functions in flat space $c_\CO$ and $f_{\phi\phi\CO}$. More precisely
\eqn{
a_\CO = \frac{f_{\phi\phi\CO}b_\CO}{c_\CO}\frac{J!}{2^J(\nu)_J}\,,
}[eq:aOdef]
$(a)_n = \frac{\Gamma(a+n)}{\Gamma(a)}$ being the Pochhammer symbol. In two dimensions ($\nu=0$) the definitions change slightly, but this will be addressed in the next subsection. The aim of this section is to compute the Fourier transform of the thermal conformal block on the Matsubara frequencies $\omega_n = 2\pi n$. Since the OPE is defined only within the unit disk, taking an integral over all space is not fully justified. However we can expect to get a good approximate answer in the limit of large spatial momenta $|\bfk|\gg 1$ where the integral will be dominated by the OPE region. Moreover, the Fourier transform will not be well defined for all possible values of $\Delta$ and $\Delta_\phi$. In such cases the answer will be obtained by analytic continuation.
\par
The integral we need to perform is
\eqn{
G^{(\Delta_\phi,d)}_{\Delta,J}(|\mathbf{k}|,\omega_n)=\int_{\BBR^{d-1}}\di^{d-1}\bfx \int_0^1 \di \tau\, e^{-i\lsp \mathbf{k}\cdot\bfx}\lsp e^{-i\lsp\omega_n \tau}\,C_J^{(\nu)}\!\left(\frac{\tau}{|x|}\right)\,|x|^{\Delta-2\Delta_\phi}\,.
}[]
First let us expand the Gegenbauer polynomial using its definition.
\eqn{
C_J^{(\nu)}(\xi) = \sum_{j=0}^{\lfloor J/2 \rfloor} c_{J,j}^{(\nu)}\,\xi^{J-2j}\,,\qquad c_{J,j}^{(\nu)} \equiv (-1)^j \frac{\Gamma(J-j+\nu)}{\Gamma(\nu)j!(J-2j)!}2^{J-2j} \,.
}[eq:gegenbauer]
For convenience let us define
\eqn{
\jt = \frac{J}{2}-j\,,\qquad \Deltat = \frac{\Delta}{2}-\Delta_\phi\,,\qquad \alpha = \jt - \Deltat\,,\qquad \beta_\pm = \frac{d}{2}+\Deltat \pm \jt\,.
}[eq:defTilde]
With these definitions the integral over $\BBR^{d-1}$ evaluates to
\eqn{
\int_{\BBR^{d-1}}\di^{d-1}\bfx\, \frac{e^{-i \bfk \cdot \bfx }}{(\bfx^2+\tau^2)^\alpha} = \frac{2^{\beta_-+\frac12}\pi^{\nu+\frac12}}{\Gamma(\alpha)}\left(\frac{\tau}{k}\right)^{\beta_--\frac12}K_{\beta_--\frac12}(k\tau)\,,
}[eq:Intdx]
where $K_\beta(z)$ is a modified Bessel function of the second kind and $k = | \bfk|$. The proof of this result is presented in \appendixname~\ref{app:integral}.
\par 
In order to perform the integral in $\tau$ we first expand the Bessel function~\cite{wolfram_besselK} around $k \to \infty$ since this is the limit where the Fourier transform can be trusted.
\eqn{
K_{\beta}(z) \underset{|z|\to \infty}{\sim} \sqrt{\frac{\pi}{2}}\frac{e^{-z}}{\sqrt{z}}\sum_{m=0}^\infty \frac{(-1)^m\left(\lifrac12+\beta\right)_m\left(\lifrac12-\beta\right)_m}{2^m\, m!}\frac{1}{z^m} \,.
}[]
And we can then perform the integral term by term in the $m$ sum
\eqn{
\int_0^1\di\tau\,e^{-(k+i\omega_n)\tau}\,\tau^{a-1} = \frac{\gamma(a,k+i\omega_n)}{(k+i\omega_n)^a}\,,
}[]
$\gamma(s,x)$ being the lower incomplete gamma function. It approaches the full $\Gamma(s)$ with exponential corrections~\cite{wolfram_incompletegamma}:
\eqn{
\gamma(a,z)\sim \Gamma(a) - e^{-z}z^{a-1}\sum_{n=0}^\infty(-1)^n(1-a)_n \frac{1}{z^n}\,.
}[eq:incompletegamma]
The result is a power series in $1 + i\omega_n/k$, plus some exponential corrections coming from the second piece of \eqref{eq:incompletegamma}. The OPE in momentum space is believed to provide an asymptotic expansion in $1/k$, see e.g.~\cite{Shifman:1978bx, Shifman:1998rb}. For this reason we expect that our approach will fail to correctly capture the non perturbative exponential corrections $\sim e^{-k}$. Therefore we will neglect all such contributions. The result is
\eqn{
G^{(\Delta_\phi,d)}_{\Delta,J}(k,\omega_n) = \sum_{j=0}^{J/2} c_{J,j}^{(\nu)} \lsp\frac{2^{\beta_-+\frac12}\pi^{\nu+\frac12}}{\Gamma(\alpha)}\sum_{m=0}^\infty \kappa_m^{(\beta_-)}\left(\frac{\Gamma\left(\beta_+-m\right)}{k^{\beta_-+m}(k+i\omega_n)^{\beta_+-m}}+O(e^{-k})\right)\,.
}[]
where $c_{J,j}^{(\nu)}$ was defined in \eqref{eq:gegenbauer} and $\kappa_m^{(\beta)}$ comes from the Bessel $K$ expansion
\eqn{
\kappa_m^{(\beta)} = \sqrt{\frac{\pi}{2}}\frac{(-1)^m\left(\beta\right)_m\left(1-\beta\right)_m}{2^m\, m!}\,.
}
The sum in $m$ can be evaluated exactly. Indeed after replacing $$\Gamma(\beta_+-m) = \frac{(-1)^m \Gamma(\beta_+)}{(1-\beta_+)_m}\,,$$
it is easy to recognize the Hypergeometric $\F$ series expansion, resulting in

\eqn{
G^{(\Delta_\phi,d)}_{\Delta,J}(k,\omega_n) = \sum_{j=0}^{J/2} c_{J,j}^{(\nu)} \lsp\frac{2^{\beta_-}\pi^{\frac{d}{2}}\Gamma\left(\beta_+\right)}{\Gamma(\alpha)}\frac{\F\left(\beta_-,1-\beta_-;1-\beta_+;\lifrac{k+i\omega_n}{2k}\right)}{k^{\beta_-}(k+i\omega_n)^{\beta_+}}+O(e^{-k})\,.
}[eq:FourierTransf]
In \sectionname~\ref{sec:analyticprop} we will study the zeros and singularities of this result.

\subsection{Two dimensions}
Two dimensional CFTs are a special case for two reasons. First and most importantly the space $\BBR\times S^1$ is conformally equivalent to $\BBR^2$, unlike the higher-dimensional analogues. As a consequence one can compute the two-point function exactly in flat space and obtain the thermal correlator by a conformal mapping. Furthermore, by Virasoro symmetry only the vacuum module $\CV_{0,0} \supset \{\unit, T_{\mu\nu},\ldots\}$ can have a non vanishing one-point function. Thus everything is known about the OPE. We will ignore for now this difference and treat $d=2$ in the same way as $d>2$. This section only deals with the following very small issue: equations \eqref{eq:euclidOPE} and \eqref{eq:aOdef} are not defined for $d=2$ ($\nu = 0$). This is a trivial problem because the dependence on $\nu$ actually simplifies. So one should write for $d=2$
\eqn{
\langle \phi(\bfx,\tau) \phi(\mathbf{0},0)\rangle_\beta = \sum_{\CO \in \phi\times\phi} a_\CO \,T_J(\xi)\,|x|^{\Delta-2\Delta_\phi}\,,
}[eq:euclidOPEd2]
\eqn{
a_\CO = \frac{f_{\phi\phi\CO}b_\CO}{c_\CO}\frac{2^{1-J}}{1+\delta_{J,0}}\,,
}[eq:aOdefd2]
where $T_n(x)$ are Chebychev polynomial of the first kind, defined as $T_n(\cos \theta) = \cos n\theta$. By following exactly the same steps as the previous section we can compute the Fourier integral, resulting in the same expression as \eqref{eq:FourierTransf} with the replacement
\eqn{
c_{J,j}^{(\nu)} \to t_{J,j} \equiv \frac{J}{2}(-1)^j\frac{(J-j-1)!}{j! (J-2j)!}2^{J-2j}\,,
}[]
coming from the coefficients in the $T_J$ polynomials. And $k \to |k|$. Of course \eqref{eq:euclidOPEd2} is useful only for obtaining the Fourier transformed block. For computations in position space it is much easier to use the thermal blocks in the $z,\zb$ coordinates. They are easy to compute even for spinning fields: the contribution of an operator $\CO$ of weights $h,\hb$ in a two-point function of a field with weights $h_\phi,\hb_\phi$ is simply
\eqn{
z^{h-2h_\phi} \,\zb^{\hb-2\hb_\phi}\,,
}
where, as usual, $z = \tau + ix,\,\zb=\tau-i x$. Note that, since $\CO \in \CV_{0,0}$, the weights $h,\hb$ are integer numbers.

\section{Analytic properties of the Fourier transformed block}\label{sec:analyticprop}

\subsection{Zeros at the double twist dimensions}

The most interesting feature of \eqref{eq:FourierTransf} are the zeros at $\alpha \in - \BBN$ coming from the $\Gamma$ function pole at the denominator. This is a feature of the analytic continuation to arbitrary values of $\Delta$ and $\Delta_\phi$: the correct result of the integral for $\alpha = 0$ would be proportional to $\delta^{d-1}(\bfk)$ and it would be infinite for negative integers. However let us define the Fourier transform as the analytic continuation from arbitrary $\Delta,\Delta_\phi$. This has interesting consequences as we can see by unpacking the definition of $\alpha$:
\eqn{
\alpha = \frac{J}{2} - j + \Delta_\phi - \frac{\Delta}{2}\in -\BBN\,.
} 
This means that if the operator has dimension
\eqn{
\Delta = 2\Delta_\phi + J + 2n\,,
}[]
for some non negative integer $n$, then $G_{\Delta,J}^{(\Delta_\phi,d)} = 0$. These are precisely the operators termed \emph{double twist operators} and can be interpreted as composite operators of the form ``$\phi\lsp \square^n \partial^{\mu_1}\cdots\partial^{\mu_J}\phi$''. Normally they appear only in free theories and acquire anomalous dimensions in general interacting theories. However there are some examples where the anomalous dimensions can be computed in perturbation theory and some operators remain exact double twist at the lowest order in the perturbation parameter. In \sectionname~\ref{sec:examples} we will discuss the generalized free theory and two of such examples to the lowest non-trivial order in their respective perturbation parameters. The appearance of these zeros is obviously not a coincidence, we will discuss its interpretations and some consequences in the next section.

\subsection{Poles for special values of the external dimension}\label{sec:dangerouspoles}

The thermal conformal blocks \eqref{eq:FourierTransf} appear to have poles at $\beta_+ \in \BBZ$. If $\beta_+$ is a non positive integer $\Gamma(\beta_+) = \infty$ and if it is positive $\F(\beta_-,1-\beta_-;1-\beta_+;z) = \infty$. However when $\beta_+$ is integer, by definition also $\beta_-$ is, thus the Hypergeometric reduces to a polynomial and the only poles that remain are the integers satisfying
\eqn{
\beta_+ \leq \max(-\beta_-,\beta_--1)\,.
}[]
Due to the unitarity bound on $\Delta$ there is only a finite number of discrete values that meet these conditions
\eqn{
\Delta = 2\Delta_\phi - d - m\,,\qquad m \leq \min(\Delta_\phi+1-d,J)\,,\qquad m \in \BBN\,.
}[]
If the spin $J$ is zero then either $\Delta_\phi = d/2$, implying a singularity in the identity operator, or $\Delta_\phi \geq (3d-2)/4$. We do not have an interpretation of these singularities. Theories that contain such operators in the spectrum must be treated as separate cases. An example would be the Ising model in $d=2$, where the energy operator $\varepsilon$ has $\Delta_\varepsilon = 1$ and $J=0$. As a consequence the block of the identity has a divergent Fourier transform, while all the other ones are well behaved. We can claim that, apart from some coincidences like this one, we do not expect to have such dangerous operators in a generic CFT. Unlike the case of the double twist operators, we cannot interpret these ones as composite operators $\phi\phi$ since $\Delta < 2\Delta_\phi$.

\section{Relation between double twist zeros and the inversion formula}\label{sec:doubletwistinversion}

\subsection{A puzzle}

In \sectionname~\ref{sec:analyticprop} we argued that blocks associated to double twist operators vanish in momentum space. This feature reminds what happens for flat space momentum conformal blocks. More precisely, as explained in \cite{Gillioz:2018mto}, the imaginary part of the block, in the limit where all momenta are lightlike, has zeros on the scaling dimensions $\Delta = 2\Delta_\phi + 2n + \ell$. The interpretation is that the imaginary part is associated to the interacting part of the correlator, and exact double twist operators correspond to free propagation. Here we need to make this statement more precise as it is not clear where the analog of the ``imaginary part'' is being taken. Moreover this property raises a puzzle: consider two theories that differ only in their double twist spectrum. The momentum space two-point function constructed from the OPE would be the same from the two theories, how is that possible?
\par
The solution is simple once one notices that the inversion formula proposed in~\cite{Iliesiu:2018fao} also has zeros at exactly the same quantum numbers. More precisely translating from their coordinates $(z,\zb)$ to $(|\mathbf{x}|=r,\tau)$ we have
\eqn{
\Disc_x\,g(r,\tau) \equiv g(r + \epsilon, \tau ) - g(r-\epsilon, \tau ) \,.
}[discdef]
The discontinuity of a single block is proportional to
\eqn{
\Disc_x\Big[C_J^{(\nu)}(\xi)|x|^{2\Deltat}\Big] \propto \sin\big[\pi\big(\Deltat-\lifrac{J}{2}+ m\big)\big]\,,
}[]
for integer $m$. This factor has the same zeros as $\Gamma(\jt-\Deltat)^{-1}$. The resolution to the puzzle is that, due to the fact that double twist blocks invert to zero (since their discontinuity vanishes), there is no combination that can be added to a correlator while keeping KMS crossing satisfied.\footnote{See footnote \ref{footnote:KMS}.} Stated in other terms, every single block addition to a correlator breaks KMS crossing, but if one is allowed to introduce double twist blocks only, there is no way of compensating such breaking. This also implies that the OPE data of double twist operators are completely redundant and can be determined by the non-double twist coefficients together with the asymptotic behavior of the ``arcs.''
\par
This correspondence suggests that we can recast the Fourier transform as an integral over a discontinuity. In the next subsection we will try to do that.

\subsection{Representation of the Fourier transform as a discontinuity}\label{sec:representationasdisc}

In order to recast the Fourier integral as an integral over the discontinuity of $g(r,\tau)$ \eqref{discdef} we first need to integrate over the $\BBR^{d-1}$ angles. The result is often called the \emph{Hankel transform}
\eqna{
\tilde{g}(k,\omega_n) &= \int_{S^{d-3}}\di\Omega\int_0^\pi\di\theta\int_0^\infty\di r\,r^{d-2}\int_0^1\di\tau\,\sin^{d-3}\theta\,e^{-i k r \cos\theta-i\omega_n\tau}g(r,\tau)=\\&=
(2\pi)^{\nu+\frac{1}{2}}\int_0^\infty \di r\int_0^1\di\tau\,r^{2\nu}\left(k r\right)^{\frac{1}{2}-\nu}J_{\nu-\frac{1}{2}}(kr)\,e^{-i\omega_n \tau}g(r,\tau)\,,
}[]
where $J_\alpha(z)$ is a Bessel function of the first kind. It has the property that $z^{-\alpha}J_\alpha(z)$ is even in $z\to-z$. Thus, when $\nu$ is integer (i.e. when $d$ is even), we can extend by parity the integral in $r$ to the entire real line. 

\begin{figure}
\centering
\includegraphics[scale=1]{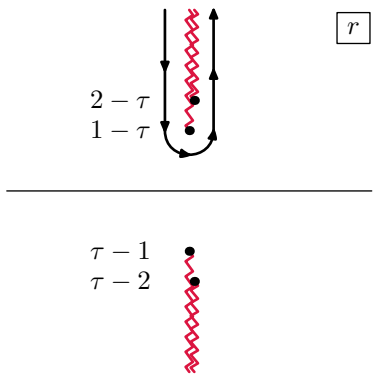}
\caption{Cuts in the complex $r$ plane and the deformed contour.}\label{fig:rplane}
\end{figure}

Let us assume from now on that $d$ is even, the case for odd $d$ is discussed in \appendixname~\ref{app:oddD}. Now we have to define the analytic continuation of $g(r,\tau)$ for complex values of $r$. Doing so will introduce branch cuts in the imaginary axis corresponding to the OPE singularities. In particular we have two series of cuts: starting from $r = i(\tau+n)$ and from $r = i(-\tau+n)$ for $n \in \BBZ$. We can choose the position of these branch cuts so that they never intersect the real axis by doing
\eqna{
\,\![r^2+(\tau-n)^2]^\delta &= [(\tau-n)+ir]^\delta [(\tau-n)-ir]^\delta\,,\qquad \tau\geq n\,,\\
&= [(n-\tau)+ir]^\delta [(n-\tau)-ir]^\delta\,,\qquad \tau< n\,.
}[]
We can then lift the contour from the real line to wrap the cuts in the upper plane as shown in \figurename~\ref{fig:rplane}. The final result is:
\eqn{
\tilde{g	}(k,\omega_n) =
(2\pi)^{\nu+\frac{1}{2}}\frac{1}{2}\int_0^{i\infty} \di r\int_0^1\di\tau\,r^{2\nu}\left(k r\right)^{\frac{1}{2}-\nu}J_{\nu-\frac{1}{2}}(kr)\,e^{-i\omega_n \tau}\,\Disc_x\,[g(r,\tau)]\,.
}[eq:FTwithdisc]
As advertised, the Fourier transform vanishes when the discontinuity is zero. The $r$ integral extends to the imaginary axis, consistently with the ``wrong'' Wick rotation performed in \cite{Iliesiu:2018fao} ($x_E \to i x_L$). However, unlike their case, here the $r$ integral extends to arbitrarily large $|r|^2$. Let us comment briefly on the integration kernel $(kr)^{\frac{1}{2}-\nu}J_{\nu-\frac{1}{2}}(kr)$. When $r$ is purely imaginary $r=is$ we can rewrite it as a modified Bessel function
\eqn{
(iks)^{-\alpha}J_{\alpha}(iks) = (ks)^{-\alpha}I_{\alpha}(ks)\,.
}[eq:BesselJandI]

\section{Spectral density}\label{sec:spectral}

\subsection{By analytic continuation to real frequency}

An important observable in Lorentzian signature is the \emph{spectral density}, defined as the Fourier transform of the commutator\footnote{The basic facts about thermal correlators are summarized in Chapter 7 of \cite{altland_simons_2010} or Chapter 2 of \cite{Meyer:2011gj}. See also \cite{le_bellac_1996,sachdev_2011}.}
\eqn{
\rho(k,\omega) = \int_{\BBR^d} \di^dx\,e^{i\omega t-i\bfk\cdot \bfx}\,\langle [\phi(x),\phi(0)] \rangle_\beta\,.
}[]
The commutator can be defined with the usual $\epsilon$-prescription, which results in the computation of a discontinuity but this time across the real time $t = -i \tau$:
\eqn{
\Disc_t\,[g(r,\tau)]\big|_{\tau=it} \equiv g(\mathbf{x},it+\epsilon) - g(\mathbf{x},it-\epsilon) = \langle[\phi(\mathbf{x},it),\phi(0,0)]\rangle_\beta\,.
}[]
An interesting property of the spectral density is that it provides an analytic continuation of the momentum space correlator $\tilde{g}(k,\omega_n)$ from the Matsubara frequencies to any complex $\omega$
\eqn{
\tilde{g}(k,\omega) = \int_{-\infty}^\infty \frac{\di \Omega}{2\pi} \frac{\rho(k,\Omega)}{i \omega + \Omega}\,.
}[]
If we let $\omega_n\to i\omega - \varepsilon$ in $\tilde{g}(k,\omega_n)$ and make use of the identity\footnote{Which is known as the Sokhotski–Plemelj theorem}
\eqn{
\lim_{\varepsilon\to 0}\frac{1}{\Omega - \omega \mp i \varepsilon} = \CP \frac{1}{\Omega-\omega}\pm i \pi \delta(\Omega -\omega)\,,
}[eq:SokhPlem]
then we can pick up the $\delta$ by taking the imaginary part to obtain
\eqn{
\rho(k,\omega) = 2\lsp\Im \, \tilde{g}(k,i\omega-\varepsilon) = -i\Disc_\omega \tilde{g}(k,i\omega)\,.
}[eq:rhoasDisc]
Having defined $\Disc_\omega f(\omega) = f(\omega+i\varepsilon)-f(\omega-i\varepsilon)$. The utility of \eqref{eq:rhoasDisc} is that, in a regime where we trust the OPE, we can express $\rho$ as a sum over primary operators
\eqn{
\rho(k,\omega) = \sum_{\CO \in \phi\times\phi} a_\CO\,\CG_{\Delta,J}(k,\omega)\,,\qquad \CG_{\Delta,J} = -i \Disc_\omega G_{\Delta,J}(k,i\omega)\,.
}[]
We drop the superscript $(\Delta_\phi,d)$ on $\CG$ and $G$ for brevity. Not only this is a representation of a real time observable in terms of Euclidean data, it also provides an inequality that can be used as a constraint on the CFT data for non-double twist operators: following from $\rho(k,\omega)>0$ if $\omega > 0$ one has
\eqn
{
-i\sum_\CO a_\CO\, \Disc_\omega\,G_{\Delta,J}(k,i\omega) > 0\,,\quad \forall \,\omega > 0\,.
}[]
Since we have already the explicit expression for $G_{\Delta,J}$, taking the discontinuity is very straightforward. However one could have defined $\CG$ by directly computing the integral with the $\epsilon$-prescription. Namely
\eqn{
\CG_{\Delta,J}(k,\omega) = \int_{\BBR^d}\di^dx\,e^{i\omega t-i\bfk\cdot \bfx}\,\Disc_t\,[C_J^{(\nu)}(\xi) \,|x|^{\Deltat}]\big|_{\tau=it} \,.
}[eq:CGasDisct]
This expression could be useful for computing the contribution to $\rho$ of those blocks that do not have a well defined Fourier transform, due to the poles discussed in \sectionname~\ref{sec:dangerouspoles}. We will not show explicitly that the two ways of computing $\CG_{\Delta,J}$ are equivalent (when both well defined), but we will prove in the next subsection that for $d=2$ they have the same support.

\subsection{The spectrum condition}\label{sec:spectrum}

\begin{figure}
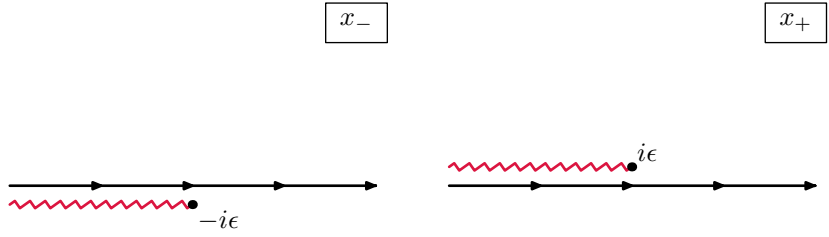

\centering
\includegraphics[scale=1]{ComplexPlane-2.mps}\qquad
\includegraphics[scale=1]{ComplexPlane-3.mps}
\caption{Cuts in the complex $x_-$ and $x_+$ plane for the first term in the definition of $\CG_{h+\hb,|h-\hb|}(k,\omega)$.}\label{fig:xPandxMPlane}
\end{figure}

Computing the discontinuity of \eqref{eq:FourierTransf} is easy. The $\omega$ dependence comes either from a power $(k-\omega)^{-\beta_+}$ or from the Hypergeometric function. The Hypergeometric $\F(a,b;c;z)$ has a discontinuity for $z>1$~\cite{wolfram_disc2F1}, which requires $\omega < -k$, where $k$ is taken to be positive. On the other hand the power has a discontinuity for $\omega > k$. All in all we get two disjoint branch cuts at $|\omega| > k$. The condition that the energy should be larger than the momentum goes under the name of \emph{spectrum condition}. As an example, the spectral density for a free theory can be computed exactly and it does satisfy the spectrum condition
\eqn{
\rho^\text{free}(k,\omega) = 2 \pi\,\mathrm{sign}\lsp(\omega)\,\delta(k^2-\omega^2)\,.
}[]
We will now show that \eqref{eq:CGasDisct} satisfies the spectrum condition too, without computing it explicitly. Let us only consider the case of $d=2$. We can go to light cone coordinates by defining
\eqn{
x_\pm=x\pm t\,,\qquad \kappa_\mp = \lifrac12(k\pm \omega)\,.
}[]
Recalling that in $d=2$ the thermal block is simply $x_+^{h-2h_\phi}x_-^{\hb-2\hb_\phi}$, the integral \eqref{eq:CGasDisct} now factorizes.
\eqn{
\CG_{h+\hb,|h-\hb|}(k,\omega) = \frac{1}{2}\int\di x_-\,\frac{e^{-i\kappa_-x_-}}{(x_-+i\epsilon)^{2h_\phi-h}} \int\di x_+\,\frac{e^{-i\kappa_+x_+}}{(x_+-i\epsilon)^{2\hb_\phi-\hb}}- (\epsilon \to -\epsilon)\,.
}[eq:CGfromepsilon]
In \figurename~\ref{fig:xPandxMPlane} are shown the planes $x_-$ and $x_+$ with the branch cuts from the first term. The second term has the cuts in the opposite half plane. If $\kappa_+$ and $\kappa_-$ have the same sign it is possible to close the contour of one integral for each of the two terms, thus by analyticity the result is zero. If on the other hand $\kappa_-$ and $\kappa_+$ have opposite signs, in one term it will not be possible to close either contour and $\CG$ will be a product of two integrals over the branch cut. In this case $\kappa_-\kappa_+  \propto k^2-\omega^2 \leq 0$, namely the spectrum condition $|\omega| > |k|$ holds.

Since we can trust the OPE representation of the spectral density when $k \gg 1$, we can make the following claim:
\begin{quote}
\emph{For any CFT at finite temperature $T$, in the limit $k \gg T$ and $y = \omega/k$ fixed, the spectral density $\rho(k,\omega)$ has support over $|y| \geq 1$, satisfying the spectrum condition.}
\end{quote}

Based on the discussions made when we obtained the Fourier transformed blocks, we believe that the corrections to this statement will be exponential $\sim e^{-k}$. However we have no rigorous way of proving this.

Recently in the literature \cite{Banerjee:2019kjh} appeared a result stating that Wightman $n$-point correlators computed at $k \gg \omega$ and $\omega$ fixed are exponentially suppressed in the spatial momenta. The case of two-point functions implies that $\rho(k,\omega) \to 0$ for $k\gg \omega$, so our result is compatible with their theorem.

\subsection{Sum rules}\label{sec:sumrules}

The spectral density has to satisfy sum rules of the form \cite{Romatschke:2009ng,CaronHuot:2009ns,Chowdhury:2016hjy,Chowdhury:2017zfu} 
\eqn{
\int \frac{\di \omega}{\pi \omega^n} \delta \rho(0,\omega) = \langle \mbox{one-point functions}\rangle\,,
}[eq:sumrule]
where $\delta\rho$ is the spectral density with some terms subtracted in order to make the integral well defined (normally one needs to subtract the $T=0$ spectral density). In \cite{Romatschke:2009ng,CaronHuot:2009ns,Chowdhury:2016hjy} $\rho$ is the spectral density for the shear viscosity correlator $\langle T_{12}(x) T_{12}(0)\rangle$, but similar (and simpler) expressions can be obtained for general scalar fields. These sum rules are defined for $k=0$, therefore it might not be possible to impose them block by block as our approximation of $k\gg1$ will break down.
\par 
Another approach for deriving sum rules may be to imitate the formalism of the deep inelastic scattering (DIS)~\cite{Komargodski:2012ek,Komargodski:2016gci}.\footnote{We thank Zohar Komargodski for suggesting this approach.} In the DIS case one considers the correlators of the CFT in a pure state $|P\rangle = \int \di x \,e^{i P\cdot x} \CO(x)|0\rangle$, but a similar analysis may be done for a thermal state. In particular, the kinematics is the same because the state $|P\rangle$ breaks the conformal group down to $O(d-1)$ and introduces a scale $P^2$, just like the thermal background. For instance
\eqna{
\langle P | \CO^{\mu_1\ldots \mu_\ell}(x)|P\rangle &\propto P_{\mu_1}\cdots P_{\mu_\ell} - (\mathrm{traces})\quad \longleftrightarrow\\
\langle  \CO^{\mu_1\ldots \mu_\ell}(x)\rangle_\beta &\propto \delta^0_{\mu_1}\cdots \delta^0_{\mu_\ell} - (\mathrm{traces})\,.
}[]
The logic is very similar, too. By doing the OPE inside the correlator one can expand $\vev{P|\phi_1\phi_2|P}$ into a sum with coefficients $a_\CO$ given by the three-point function on the vacuum $f_{\phi_1\phi_2\CO}$ and the one-point function of $\CO$ on $|P\rangle$. There is however an important difference between the DIS setup and the thermal correlator as we will now explain. Let us consider the kinematic variable $z = -2\, q\cdot P / q^2$ for $q^\mu = (\omega , k)$ a space-like four momentum. The DIS amplitude $\CA$ defined as
\eqn{
\CA(q^2,z) \equiv \int \di^d x\, e^{iq\cdot x} \,\langle P | \phi_1(x)\phi_2(0)| P\rangle\,,
}[]
has a discontinuity only for $z \in [-1,1]$. This allows us to take a contour around the interval $[-1,1]$ and deform it to a circle at infinity. Therefore one can write a series of sum rules
\eqn{
\int_{-1}^1\di z\,z^{s-1} \,\Im\lsp\CA(q^2,z) = \frac1{2i}\oint_\infty\di z\,z^{s-1}\lsp \CA(q^2,z)\,,
}[]
for any sufficiently large $s$.\footnote{Assuming that $\Im\lsp\CA$ is polynomially bounded by some $z^{-N+1}$ as $z \to 0$ this requires $s\geq N$.} We are not allowed to make a similar argument in our case because $\rho(k,\omega)$ is in general nonzero for all $\omega \in \BBR$. As a consequence we cannot find a contour that completely encircles the discontinuity.
\par
However this problem disappears in the $k\gg 1$ limit because, thanks to the result obtained in the previous subsection, $\rho(k,\omega)$ satisfies the spectrum condition. We can define the variable $z = k/\omega$ so that the discontinuity of $\tilde{g}(k,i\omega)$ is nonzero only when $z\in [-1,1]$. By following the same steps we find the following sum rules for $k\gg 1$
\eqn{
\int_{-1}^1\di z\,z^{s-1}\,\rho(k,k/z) = \frac1i\oint_\infty\di z\,z^{s-1}\,\tilde{g}(k,ik/z)\,,
}[eq:sumrulethermalDIS]
for any sufficiently large integer $s$. Assuming the OPE is convergent for large $k$, the right hand side can be computed block by block. Sadly, however, there is no simple closed form expression. Letting $y = 1/z$
\eqn{
\mathrm{r.h.s.}\eqref{eq:sumrulethermalDIS} = \sum_\CO a_\CO \Bigg[
\sum_{j=0}^{J/2} c_{J,j}^{(\nu)} \frac{2^{\beta_-}\pi^{\frac{d}2}\Gamma(\beta_+)}{\Gamma(\alpha)k^{d+2\tilde{\Delta}}}\,
i\oint_0 \frac{\di y}{y^{s+1}}\,\frac{\F\big(\beta_-,1-\beta_-;1-\beta_+;\tfrac{1-y}2\big)}{(1-y)^{\beta_+}}
\Bigg]\,.
}[]
The integral can be easily evaluated by the theorem of residues because the poles at $y=0$ are coming only from the $y^{-s-1}$ factor. Notice however that, unlike the DIS case, the integral is nonzero even for operators with spin different from $s$.
\par
It remains to motivate that the integral in the left hand side is well defined for $z\to0$. This requires the knowledge of the behavior of $\rho(k,\omega)$ for $\omega \gg k \gg 1$. One could simply argue by dimensional analysis that $\rho(k,\omega)$ must be bounded by $|\omega|^{2\Delta_\phi-d}$ because we expect scale invariance to be restored at very high energies and therefore $\langle \phi(x)\phi(0)\rangle \sim |x|^{-2\Delta_\phi}$. Another way of saying this is that the one-point function of the identity operator dominates~\cite{Meyer:2010ii,Chowdhury:2016hjy}. There are many examples in QCD~\cite{Meyer:2008gt,Meyer:2011gj} and holography~\cite{Son:2002sd} where this is confirmed.\footnote{In those examples the operator in the correlator was the stress tensor $T$ in four dimensions rather than $\phi$. The bound becomes $2\Delta_\phi-d = 4$.}${}^{,}$\footnote{Also our example in \sectionname~\ref{sec:spectral2d} agrees with this claim.} If that is the case, any $s > 2\Delta_\phi - d$ will make the integral convergent.

\begin{figure}
\centering
\includegraphics{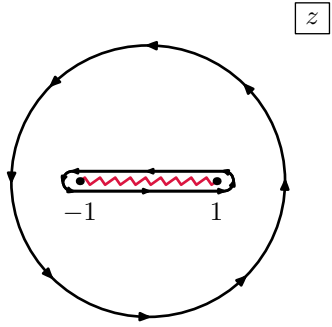}
\caption{Contour deformation in the $z$ plane of $\tilde{g}(k,ik/z)$ that leads to the sum rule \eqref{eq:sumrulethermalDIS}.}
\end{figure}

\section{Examples in general dimensions}\label{sec:examples}

\subsection{Generalized free theory}

We can use the coefficients of the generalized free theory (GFT) and test the OPE in momentum space. The coefficients $a_\CO$ were computed in \cite{Iliesiu:2018fao}.
\eqna{
a_\unit &= 1 \,,\\
a_{[\phi\phi]_{n,\ell}} &= 2 \lsp\zeta(2\Delta_\phi+2n+\ell) \frac{(\ell+\nu)(\Delta_\phi)_{\ell+n}(\Delta_\phi-\nu)_n}{n!(\nu)_{\ell+n+1}}\,.
}[]
The normalization of the two-point function is the following:
\eqn{
\langle \phi(\bfx,\tau) \phi(\mathbf{0},0)\rangle_\beta \equiv g(\bfx,\tau) =  \sum_{m\in\BBZ}\frac{1}{((\tau+m)^2+\bfx^2)^{\Delta_\phi}}\,.
}[]
We can easily compute the Fourier transform over the Matsubara frequencies:
\eqna{
\tilde{g}(\mathbf{k},\omega_n) &\equiv \int_{\BBR^{d-1}} \di^{d-1}\bfx \int_0^1 \di \tau \,e^{i \llsp\mathbf{k}\cdot \bfx}e^{i\llsp\omega_n \tau}\, g(\bfx,\tau) =\\&=
\sum_{m\in\BBZ}\int_{\BBR^{d-1}} \di^{d-1}\bfx \int_m^{m+1} \di \tau' \,e^{i \llsp\mathbf{k}\cdot \bfx}e^{i\llsp\omega_n \tau - i\llsp \omega_n m}\, \frac{1}{({\tau'}^2+\bfx^2)^{\Delta_\phi}} = \\&=
\int_{\BBR^{d}} \di^{d}x \,e^{i \llsp k\cdot x}\, \frac{1}{|x|^{2\Delta_\phi}} = \\&=
\frac{\pi^{\frac{d}{2}}\lsp\Gamma\left(\frac{d}{2}-\Delta_\phi\right)}{2^{2\Delta_\phi-d}\lsp\Gamma(\Delta_\phi)}\big(\mathbf{k}^2+\omega_n^2\big)^{\Delta_\phi-\frac{d}{2}}\,.
}[]
As discussed in \sectionname~\ref{sec:analyticprop}, \eqref{eq:FourierTransf} has zeros at the double twist dimensions. This means that every single term but the contribution from the identity vanishes! It is now trivial to show that this single term actually matches the full correlator
\eqn{
\tilde{g}(k,\omega_n) = G^{(\Delta_\phi,d)}_{0,0}(k,\omega)\,.
}[]
So in the particular case of GFT the OPE in momentum space captures the entire function exactly with only one block.

\subsection{ \texorpdfstring{$O(N)$}{O(N)} sigma model at large  \texorpdfstring{$N$}{N}}

The $O(N)$ model in odd dimension $d$ has three kinds of singlet operators at $N = \infty$.
\benum
\item The conserved higher spin currents (strictly conserved only for $N=\infty$),
\[
J_\ell = [\phi_i\phi_i]_{0,\ell}\,,\quad \Delta_{J_\ell} = \ell + d-2\,,\quad \ell > 0\,.
\]
\item The double trace operators $[\phi\phi]_{n,\ell}$, with $n>0$ and $\ell >0$,
\[
[\phi_i\phi_i]_{n,\ell}\,,\quad \Delta_{n,\ell} = \ell + 2n + d-2\,,\quad n\in \BBN^+\,.
\]
\item Powers of $\sigma$, the ``shadow'' of $\phi^2$
\[
\sigma^m \neq [\phi_i\phi_i]_{0,0}\,,\quad \Delta_m = 2m\,,\quad m\in \BBN\,.
\]
\eenum
By the same reason as in GFT, only the operators $\sigma^m$ will contribute to the sum, because the others have double twist dimensions and their blocks vanish. We better reproduce the full two-point function just by summing the tower of $\sigma^m$ then.\footnote{To be precise the actual operators for $m > 2$ might not just be $\sigma^m$ but rather a mixture of the double twists and $\sigma^m$, due to the equations of motion $\partial^2\phi_i = \sigma \phi_i$. But this will not make any difference for the following.}
\par
From \cite{Petkou:2018ynm} we can compute the OPE coefficient of $\sigma^m$
\eqn{
a_{\sigma^m} = - \frac{\sqrt{\pi/2}}{(2\pi)^\frac{d}{2}}\,\underset{\Delta = 2m}{\mathrm{Res}} \,\frac{m_{\mathrm{th}}^\Delta}{2^{\Delta - \frac{d-5}{2}}\sqrt{\pi}}\Gamma\lnsp\left(-\frac{\Delta}{2}\right)\Gamma\lnsp\left(-\frac{\Delta-d+2}{2}\right)
}[asigmam]
the factor in front is just a normalization coming from the conventions used there. The residues at the odd integers cancel when the gap equation is satisfied, as explained in the paper. A straightforward computation shows
\eqn{
a_{\sigma^m} = \frac{2^{\frac{d-3}{2}}m_{\mathrm{th}}^{2 m}\, \Gamma\lnsp\left(\frac{d}{2}-m-1\right)}{
   \Gamma\lnsp\left(\frac{1}{2}-m\right) \Gamma(2 m+1)}\,.
}[]
Now we plug this into \eqref{eq:FourierTransf} for $J = 0,\,\Delta_\phi = \frac{d-2}{2},\,\Delta = 2m$. Before doing so we notice that the Hypergeometric for $\tilde{\jmath} = 0$ can be simplified by means of
\eqn{
\F(a,b;b;z) = (1-z)^{-a}\,.
}[]
After some algebra we arrive to
\eqna{
a_{\sigma^m} G^{(\nu,d)}_{2m,0}(k,\omega_n) &=
\frac{2^{2m}m_{\mathrm{th}}^{2m}\sqrt{\pi}\lsp\Gamma(m+1)}{\Gamma(2m+1)\Gamma\big(\frac{1}{2}-m\big)}\frac{1}{(k^2+\omega_n^2)^{m+1}}\,.
}[]
We need to show then that the full correlator is reproduced, namely we need to verify this sum
\eqn{
\frac{1}{k^2+\omega_n^2+m_\mathrm{th}^2} = \sum_{m=0}^\infty
\frac{2^{2m}m_{\mathrm{th}}^{2m}\sqrt{\pi}\lsp\Gamma(m+1)}{\Gamma(2m+1)\Gamma\big(\frac{1}{2}-m\big)}\frac{1}{(k^2+\omega_n^2)^{m+1}}\,.
}[ON]
Amazingly all the $\Gamma$ functions simplify thanks to
\eqn{
\Gamma(2m + 1) = \frac{2^{2m}}{\sqrt\pi}\Gamma\big(m+\lifrac{1}{2}\big)\Gamma(m+1)\,,\quad
\Gamma\big(m+\lifrac{1}{2}\big)\Gamma\big(\lifrac{1}{2}-m\big) = (-1)^{m}\lsp\pi\,.
}[]
We obtain then a geometric series and \ON is reproduced exactly. Sadly, however, there is no apparent constraint to the value of $m_\mathrm{th}$. Any real number seems consistent with this computation. This is again a consequence of the zeros of the blocks: the value of $m_\mathrm{th}$ can be determined by requiring the absence of the spurious operators at dimensions $\Delta_{``\phi\square^{n}\phi\text{''}} = d-2+2n$, but even if they were present in the theory, they would not enter in the sum.
\par
We can also check in this particular example that taking the discontinuity block by block reproduces the correct value for the spectral density. In this case the sum is over scalar operators of dimension $\Delta = 2m$. 
\eqn{
\frac{1}{k^2 -\omega^2 + m_\mathrm{th}^2} =\sum_{m=0}^\infty a_{\sigma^{m}} G^{(\nu,d)}_{2m,0}(k,i\omega	) =  \sum_{m=0}^\infty \frac{(-m_\mathrm{th}^2)^m}{(k^2-\omega^2)^{m+1}}\,.
}[commuteDiscSum]
The discontinuities are computed thanks to a generalization of \eqref{eq:SokhPlem}. Consider the function $(k^2-\omega^2)^{-m-1}$ as a meromorphic function of $\omega^2\equiv \zeta$. The discontinuity is
\eqn{
\Disc_\omega \frac{1}{(k^2-\zeta)^{m+1}} = \frac{1}{(k^2-\zeta-i\varepsilon)^{m+1}} - \frac{1}{(k^2-\zeta+i\varepsilon)^{m+1}} \equiv \delta_{m}(\zeta)\,.
}[]
Suppose we integrate the function $\delta_m$ defined above against an arbitrary holomorphic function $f$ along the real line. In each of the two terms we can do an infinitesimal change of variables to shift both poles to $\zeta = k^2$. This will modify the integration contours in the two terms, but by holomorphicity we can deform it to a circle enclosing $\zeta = k^2$:
\eqna{
\int_{-\infty}^\infty \di \zeta\, \delta_m(\zeta)f(\zeta) &= \int_{-\infty+i\varepsilon}^{\infty+i\varepsilon} \di \zeta\frac{f(\zeta-i\varepsilon)}{(k^2-\zeta)^{m+1}} - \int_{-\infty-i\varepsilon}^{\infty-i\varepsilon} \di \zeta\frac{f(\zeta+i\varepsilon)}{(k^2-\zeta)^{m+1}}=\\&=
-\oint_{|\zeta-k^2|=\varepsilon} \frac{f(\zeta)}{(k^2-\zeta)^{m+1}} + O(\varepsilon)=\\&=
-2\pi i \,\mathrm{Res}_{\zeta\to k^2} \frac{f(\zeta)}{(k^2-\zeta)^{m+1}}+ O(\varepsilon)\,.
 }[]
Taking $\varepsilon\to 0$ we conclude that
\eqn{
\delta_m(\zeta) = \frac{2\pi i}{m!}\frac{d^m}{d\zeta^m} \delta(\zeta-k^2)\,.
}[]
Then by taking the discontinuity of \commuteDiscSum and commuting it with the sum we find
\eqn{
2\pi i\sum_{m=0}^\infty \frac{(-m_\mathrm{th}^2)^m}{m!}\frac{d^m}{(d\omega^2)^m} \delta(\omega^2-k^2) = 2\pi i \, e^{-m_\mathrm{th}^2 \frac{d}{d\omega^2}}\delta(\omega^2-k^2)\,.
}[]
In the last step we have exponentiated the series of derivatives in order to turn it into a translation operator, thus proving the desired identity.
\eqn{
\rho(k,\omega)\equiv -i \Disc \frac{1}{k^2-\omega^2+m_{\mathrm{th}}^2} = 2\pi \, \delta(\omega^2-k^2-m_\mathrm{th}^2)\,.
}[]

\subsection{First order $\epsilon$ expansion}

Let us summarize the results for the $\epsilon$-expansion for the Wilson-Fisher fixed point in $d = 4 - \epsilon$. We call $J_\ell$ (for $\ell >0$) the broken higher spin currents $J_\ell \sim \phi \lsp \partial_{\mu_1} \cdots \partial_{\mu_\ell} \phi$  and $\phi^2$ the leading twist scalar. Their dimensions are given by ($\Delta_{\phi^2}$ is related to the $\nu$ critical exponent)
\eqna{
\Delta_{\phi^2} &= d - 2 + \frac{\epsilon}{3} +O(\epsilon^2)\,,\quad\mbox{\cite{Henriksson:2018myn,Kleinert:1991rg}} \\
\Delta_{\ell} &= d-2+\ell + O(\epsilon^2)\,,\quad\mbox{\cite{Gliozzi:2017gzh,Alday:2017zzv,Roumpedakis:2016qcg}}\,.
}[deltas]
The fundamental field has an anomalous dimension (the $\eta$ critical exponent) only at second order
\eqn{
\Delta_\phi = \frac{d-2}{2} + O(\epsilon^2)\,,\quad\mbox{\cite{Kleinert:1991rg,Gliozzi:2017gzh,Alday:2017zzv}}\,.
}[]
The free theory OPE coefficients normalized as \cite{Henriksson:2018myn,Fitzpatrick:2012yx,Alday:2016jfr} read
\eqn{
f_\ell^{\text{free}} = (1+(-1)^\ell)\frac{\Gamma\lnsp\left(\lifrac{d}{2}+\ell-1\right)^2\Gamma(d+\ell-3)}{\Gamma\lnsp\left(\lifrac{d}{2}-1\right)^2\Gamma(\ell+2)\lsp\Gamma(d+2\ell-3)}\,.
}[]
The corrections in $\epsilon$ will not be needed to this order. These are four-point conformal block coefficients. In order to relate them to squares of OPE coefficients we should use the normalization in \cite{Poland:2018epd}, namely
\eqn{
\lambda_{\phi\phi J_\ell}^2 = (-2)^\ell f_\ell\,,\quad \lambda_{\phi\phi \llsp\phi^2}^2 = f_0\,.
}[]
Lastly, we will need the one-point function coefficients. Also in this case the free theory value is sufficient.
\eqn{
b_\ell = \frac{2^{\ell+1}}{\lambda_{\phi\phi J_\ell}}\lsp\zeta(2+\ell)\,,\qquad b_0 = \frac{2}{\lambda_{\phi\phi  \llsp\phi^2}}\lsp\zeta(2)\,,\quad \mbox{\cite{Iliesiu:2018fao}} \,.
}[]
As a reminder, the coefficient $a_\CO$ appearing in the thermal OPE is defined as
\eqn{
a_\ell =  \frac{\ell!}{2^\ell\left(\frac{d-2}{2}\right)_\ell}\lsp \lambda_{\phi\phi J_\ell} b_\ell \,,\qquad
a_0 =\lambda_{\phi\phi \llsp\phi^2} b_0\,.
}[as]
\par
These operators are sufficient for describing the theory up to order $\epsilon$. At order $\epsilon$ all blocks but the identity and $\phi^2$ vanish. The block of $\phi^2$ is already of order $\epsilon$ thus we will only need the OPE coefficient at tree level. To order $\epsilon^2$ we have all double twist operators appearing with their tree level OPE coefficient and the operator $\phi^2$ with his second order anomalous dimension or with his first order OPE coefficient. Operators with non leading twist like $\phi \partial^k \phi \partial^l \phi$ start appearing from order $\epsilon^4$~\cite{Henriksson:2018myn}. 
\par

Following the result of \cite{Thesis:Yang}, the correlator at one loop is simply a free propagator with a thermal mass given by $\lambda T^2$, $T$ being the temperature (here set to 1) and $\lambda$ the quartic coupling $\mathcal{L}_{\mathrm{int}} = \lambda \lsp \phi^4$. The value of the coupling at the critical point is \cite{Kleinert:1991rg}
\eqn{
\lambda = \frac{16\pi^2}{4!}\frac{\epsilon}{3}
+O(\epsilon^2)\,.}[]
And the correlator reads
\eqn{
g(k,\omega_n) = \frac{1}{\omega_n^2 + k^2 + \lambda}\,.
}[corroneloop]
If one normalizes the two-point function in this way, the blocks need to be rescaled by a normalization $\mathcal{N}_d$
\eqn{
\mathcal{N}_d = \frac{\Gamma\left(\frac{d-2}{2}\right)}{4\lsp\pi^{d/2}}\,.
}[]
\par
As noticed before, at order $\epsilon$ we have only the contribution of the identity and of $\phi^2$. The momentum space thermal block in full generality is in \eqref{eq:FourierTransf}. It is easy to compute the scalar block to first order in the anomalous dimension
\eqn{
G_{2-\epsilon + \epsilon \lsp \gamma_0,0}^{(\frac{2-\epsilon}{2},4-\epsilon)}(k,\omega_n) = - \frac{2\llsp \epsilon\lsp \gamma_0}{(k^2 + \omega^2_n)^2}\,,
}[]
where in this case $\gamma_0$ is 1/3 \deltas. The one loop correlator in \corroneloop is expanded to
\eqn{
g(k,\omega_n) = \frac{1}{\omega_n^2 + k^2} - \frac{2\llsp \pi^2 \epsilon}{9(\omega_n^2 + k^2)^2}\,.
}[]
And it can be easily checked that, to order $\epsilon$, this is equal to
\eqn{
g(k,\omega_n)  = \CN_{4-\epsilon} \left( G_{0,0}^{(\Delta_{\phi},4-\epsilon)}(k,\omega_n) + a_0\lsp G_{\Delta_{\phi^2},0}^{(\Delta_{\phi},4-\epsilon)}(k,\omega_n)\right)\,,
}[]
where $a_0$ was defined in \as and it is simply $2\llsp\zeta(2) = \pi^2/3$. Thus the OPE (which is made by two terms in this case) reproduces the full answer to order $\epsilon$.

\section{Example in two dimensions}\label{sec:ex2d}

As previously mentioned, two dimensional CFTs are special because the thermal cylinder $\BBR \times S^1$ is conformally equivalent to the plane. The explicit conformal map is (we will restore the explicit definition of $\beta$ only for this formula)
\eqn{
z \to z(w) =\exp\left(\frac{2\pi w}{\beta}\right)\,.
}[]
So that if $w = r + i \tau$, then $\tau$ is periodic with period $\beta$. The two-point function in flat space reads
\eqn{
\langle \phi_{h_\phi,\hb_\phi}(z,\zb)\,\phi_{h_\phi,\hb_\phi}(0,0)\rangle = \frac{1}{z^{2h_\phi}\lsp \zb^{2\hb_\phi}}\,.
}[]
Thus the conformal transformation $z(w)$ maps it to
\eqn{
\langle \phi_{h_\phi,\hb_\phi}(w,\wb)\,\phi_{h_\phi,\hb_\phi}(0,0)\rangle = \frac{\pi^{2(h_\phi+\hb_\phi)}}{\sinh^{2h_\phi}(\pi w)\lsp \sinh^{2\hb_\phi}(\pi \wb)}\,.
}[eq:twopftwod]
In \appendixname~\ref{app:coeffaO} we show how to compute the coefficients $a_\CO$ from the OPE and check that they match the Taylor expansion of \eqref{eq:twopftwod}.

\subsection{The spectral density from the exact correlator}\label{sec:spectral2d}

The spectral density obtained from \eqref{eq:twopftwod} reads
\eqn{
\rho(k,\omega) =  \int_{\BBR^2} \di^2x\,e^{i\omega t-i k x}\,
 \frac{\pi^{2(h_\phi+\hb_\phi)}}{\sinh^{2h_\phi}[\pi( x-t+i\epsilon)]\lsp \sinh^{2\hb_\phi}[\pi( x+t-i\epsilon)]}
- (\epsilon \to -\epsilon)
}[]
It can be computed by factorizing the integral. Defining as before
\eqn{
x_\pm = x \pm t\,,\qquad \kappa_\mp = \lifrac12(k\pm \omega)\,,
}[] 
one has
\eqn{
\rho(k,\omega) = \frac{1}{2}\int\di x_-\,\frac{\pi^{2h_\phi}\,e^{-i\kappa_-x_-}}{\sinh^{2h_\phi}[\pi( x_-+i\epsilon)]} \int\di x_+\,\frac{\pi^{2\hb_\phi}\,e^{-i\kappa_+x_+}}{\sinh^{2\hb_\phi}[\pi( x_+-i\epsilon)]}- (\epsilon \to -\epsilon)\,.
}[]
Let us define
\eqn{
\CH^{\pm}_h(\kappa) \equiv \int\di x\,\frac{\pi^{2h}\,e^{-i\kappa x}}{\sinh^{2h}[\pi( x\pm i\epsilon)]}\,.
}[]
The integral for general $h$ was computed in \cite{Becker:2014jla,Maldacena:1997ih} and the result is
\eqn{
\CH^{\pm}_h(\kappa) = e^{\mp \pi h} \frac{(2\pi)^{2h-1}}{\Gamma(2h)}\, e^{\pm \kappa/2}\, \Gamma\lnsp\left(h+ \frac{i\kappa}{2\pi}\right)\llsp \Gamma\lnsp\left(h- \frac{i\kappa}{2\pi}\right)\,.
}[]
For the example we plan to show we need this integral for $h=1/2$. It can be obtained from the formula above by using the property $\Gamma(z)\Gamma(1-z)=\pi/\sin(\pi z)$
\eqn{
\CH_{1/2}^{\pm}(\kappa) = \mp i \frac{\pi\lsp e^{\pm\kappa/2}}{\cosh (\kappa/2)}\,.
}[]
Then the spectral density is readily computed as
\eqn{
\rho(k,\omega) = \frac{1}{2}\big(\CH_{1/2}^{+}(\kappa_-)\CH_{1/2}^{-}(\kappa_+) - \CH_{1/2}^{-}(\kappa_-)\CH_{1/2}^{+}(\kappa_+) \big) = \frac{2\pi^2\, \sinh (\omega/2)}{\cosh (k/2) + \cosh(\omega/2)}\,.
}[eq:spectralh12]

It is interesting to consider the limit for $k,\omega \gg 1$ with ratio fixed $y = \omega/k$. If we have $|y| < 1$ then \eqref{eq:spectralh12} goes to zero whereas if $|y| > 1$ it asymptotes to $\pm 2\pi^2$.
\eqn{
\rho(k,\omega) \underset{\substack{k \gg 1\,,\\y\llsp=\llsp\omega/k}}{\sim} 2\pi^2\,\Theta(|y|-1) \,\mathrm{sign}\lsp(\omega)\,.
}[eq:limitrhotwod]
Moreover the corrections are exponentially small (i.e. non perturbative in $1/k$). This agrees with our general claim in \sectionname~\ref{sec:spectrum}. In the next subsection we will show that the discontinuity of the thermal blocks reproduces correctly this limit.
\par
Due to its relevance for the sum rules defined in \eqref{eq:sumrulethermalDIS} we will also comment on the limit $\omega \gg k \gg 1$ for general $h_\phi,\hb_\phi$. Using the asymptotic expansion of $\Gamma$ functions for large imaginary part~\cite{wolfram_gammalargeim}
\eqn{
\vert\Gamma(a + i b)\vert^2 \underset{|b|\to \infty}{\sim} 2\pi |b|^{2a-1}e^{-\pi |b| -2a}\,,
}[]
it is easy to show that
\eqn{
\rho(k,\omega) \underset{\omega \gg k \gg 1}{\sim} C_{h_\phi,\hb_\phi}\, |\omega|^{2(h_\phi + \hb_\phi)-2}\,,
}[]
for some constant $C_{h_\phi,\hb_\phi}$. This agrees with the general bound $|\omega|^{2\Delta_\phi - d}$ explained in \sectionname~\ref{sec:sumrules}.

\subsection{Spectral density from the blocks}

The computation of the spectral density from the OPE is very simple as we will see. First we recall the issue stated at \sectionname~\ref{sec:dangerouspoles}: for this particular correlator with $h_\phi = \hb_\phi = 1/2$ the block of the identity is not well defined. All other blocks however are well defined and they read
\eqn{
G^{(1,2)}_{h+\hb,|h-\hb|}(k,\omega_n) = \sum_{j=0}^{|h-\hb|/2} t_{J,j} \lsp\frac{2^{\hb+j}\pi\Gamma\left(h-j\right)}{\Gamma(1-\hb-j)}\frac{\F\left(\hb+j,1-\hb-j;1-h+j;\lifrac{k+i\omega_n}{2k}\right)}{k^{\hb+j}(k+i\omega_n)^{h-j}}\,,
}[]
if $h>\hb$, otherwise one has to flip $h\leftrightarrow \hb$. It can be immediately seen that all discontinuities manifestly vanish as all powers are integer and for every non negative value of $\hb$ the Hypergeometric reduces to a polynomial.\footnote{For it to be an infinite series one must have $-j< \hb < -j+1$, which is impossible for integer $\hb$.} Let us now use the alternative way presented in equation \eqref{eq:CGfromepsilon} to compute the discontinuity of the identity block and to check this result for the other blocks.

\begin{figure}
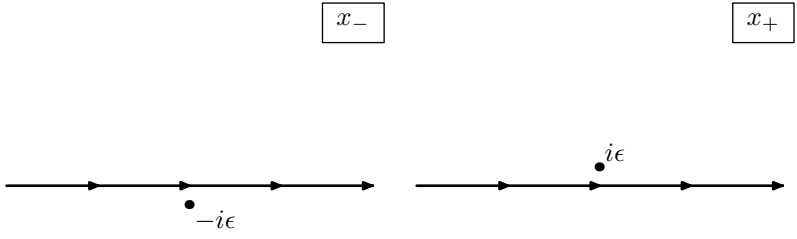

\centering
\includegraphics[scale=1]{ComplexPlane-4.mps}\quad
\includegraphics[scale=1]{ComplexPlane-5.mps}
\caption{Poles in the complex $x_-$ and $x_+$ planes for the computation of the identity discontinuity.}\label{fig:residuesidentity}
\end{figure}

\eqn{
\CG_{h+\hb,|h-\hb|}(k,\omega) = \frac{1}{2}\int\di x_-\,\frac{e^{-i\kappa_-x_-}}{(x_-+i\epsilon)^{1-h}} \int\di x_+\,\frac{e^{-i\kappa_+x_+}}{(x_+-i\epsilon)^{1-\hb}}- (\epsilon \to -\epsilon)\,.
}[]
As one can immediately see, for either $h$ or $\hb$ greater than zero, at least one of the two integrands is analytic and therefore the contour can be closed and $\CG$ evaluates to zero. If $h=\hb = 0$ as a straightforward application of the theorem of residues we obtain (see \figurename~\ref{fig:residuesidentity})
\eqn{
\CG_{0,0}(k,\omega) = 2\pi^2\,\Theta(-\kappa_+\kappa_-)\,\mathrm{sign}\lsp(\kappa_-) =  2\pi^2\,\Theta(\omega^2-k^2) \,\mathrm{sign}\lsp(\omega)\,.
}[]
in agreement with the limit in \eqref{eq:limitrhotwod}.

\section{Outlook}
We computed the Fourier transform of the thermal conformal blocks and observed that in several examples in any $d$ the OPE exactly reproduces the momentum space Green's function. It would be interesting to better understand the limitations of this approach, for example considering $\epsilon$ expansion or large $N$ expansion to the next order or maybe theories where many CFT data are known numerically---like the 3$d$ Ising model \cite{Simmons-Duffin:2016wlq,Iliesiu:2018zlz}.
\par
It would be of great theoretical significance to understand and perhaps quantify the corrections to the OPE approximation in momentum space. This would be useful for estimating the deviations from our theorem about the spectral density: $\lim_{k\to \infty}\rho \propto \Theta(|\omega/k|-1)$.
\par
Finally, it would be interesting to systematically investigate all constraints on the CFT data stemming from \eqref{eq:sumrule}. They suffer from being defined at $k=0$, which is the opposite limit of the one considered in this formalism. However since in many examples that we studied the block expansion worked all the way down to $k=0$, there is reason to believe that it can work for other cases too. The DIS sum rules \eqref{eq:sumrulethermalDIS} on the other hand are well defined within our formalism and are especially promising for theories with many double twist operators, where the right hand side of the sum rule becomes much simpler.

\ack{I am very grateful to Zohar Komargodski for suggesting this project and providing many helpful ideas and comments during its realization. I would like to thank An­dreas Ster­giou and Dalimil Maz\'a\v{c} for interesting discussions in the early stages of this work. I would also like to thank the Simons Center for Geometry and Physics for its hospitality throughout the entire preparation of this manuscript. I am supported by the Swiss National Science Foundation under grant no.\ PP00P2-163670 and by the Doc.Mobility program of the Swiss National Science Foundation.}

\Appendices

\section{Proof of the integral}\label{app:integral}
We prove formula \eqref{eq:Intdx} by induction. The integral can be evaluated in $d=2,3$ and $4$ explicitly, showing agreement with the formula. The case of $d=2$ is an integral representation of the Bessel function~\cite{wolfram_besselKint}. For $d=4$ the integral over the angle is straightforward and, after an integration by parts, it is possible to use \cite{wolfram_besselKint} again. Lastly, if $d=3$ the integral over the angle yields a Bessel function $J_0(k r)$ and the final result is given by the Hankel transform of $(\tau^2+r^2)^{-\alpha}$ of order zero~\cite{PoularikasHandbook}. For $d>2$ let us write
\eqn{
\CI_d(k,\alpha) \equiv \int_0^\infty\di r\int_0^\pi\di\theta\,r^{d-2}\sin^{d-3}\theta \frac{e^{-i\lsp kr\lsp \cos\theta}}{(r^2+\tau^2)^\alpha}\,.
}[]
The above expression can be easily seen to satisfy a recurrence equation by differentiating twice with respect to $k$
\eqn{
\frac{\di^2}{\di^2 k}\mathcal{I}_d(k,\alpha) = \mathcal{I}_{d+2}(k,\alpha) - \mathcal{I}_{d}(k,\alpha-1)+\tau^2 \lsp\mathcal{I}_d(k,\alpha)\,.
}[eq:recrel]
The integral we are after is $\frac{2\pi^\nu}{\Gamma(\nu)}\CI_d(k,\alpha)$. The result in \eqref{eq:Intdx} can be shown to satisfy the recursion relation \eqref{eq:recrel}, therefore by induction the formula is proven.

\section{Fourier transform as a discontinuity in odd dimensions}\label{app:oddD}

Here we want to show that for odd $d$ we can still write an integral expression for $\tilde{g}(k,\omega_n)$ whose integrand is proportional to $\Disc_x[g(r,\tau)]$, similar to \eqref{eq:FTwithdisc}. Naturally we can always split the integral into a $d-1$ dimensional one and an integral over $\BBR$.
\eqna{
\tilde{g}(k,\omega_n) &= (2\pi)^{\nu'+\frac{1}{2}}\int_{-\infty}^\infty\di x^d\,e^{-i k^d x^d}\int_0^\infty \di r\int_0^1\di\tau\\
&\qquad r^{2\nu'}\left(k r\right)^{\frac{1}{2}-\nu'}J_{\nu'-\frac{1}{2}}(kr)\,e^{-i\omega_n \tau}g\Big(\sqrt{r^2+(x^d)^2},\tau\Big)\,,
}[]
where $\nu$ is replaced by $\nu' = (d-3)/2$, which is integer. Now one can run the same argument that led to \eqref{eq:FTwithdisc} in Sec.~\ref{sec:representationasdisc} and obtain an integrand over the discontinuities, which are now shifted to $r = i(\tau \pm n \pm x^d)$
\eqna{
\tilde{g}(k,\omega_n) &= (2\pi)^{\nu'+\frac{1}{2}}\int_{0}^\infty\di x^d\,e^{-i k^d x^d}\left[\int_{ix^d}^{i\infty} \di r\int_0^1\di\tau\right.\\
&\qquad r^{2\nu'}\left(k r\right)^{\frac{1}{2}-\nu'}J_{\nu'-\frac{1}{2}}(kr)\,e^{-i\omega_n \tau}\Disc_x[g(r,\tau)] + (x^d \to -x^d)\bigg]\,.
}[]
We surely get a more involved integral expression as we have to further integrate in $x^d$. But we have obtained a representation that explicitly shows an integrand proportional to $\Disc_x[g(r,\tau)]$ as needed.

\section{Compute the $a_\CO$ starting from one and three point functions}\label{app:coeffaO}

In flat space all one-point functions vanish, except for the identity. Therefore in the cylinder the only one-point functions that can appear are those of the operators inside the vacuum Verma module $\CV_{0,0}$. That is, all Virasoro descendants $L_{-Y}\Lb_{-\Yb}\unit$, where $Y = \{k_1,\ldots,k_m\}$ is a multi-index 
\eqn{
L_{-Y}\Lb_{-\Yb} \equiv L_{-k_1}\cdots L_{-k_m}\,\Lb_{-\kb_1}\cdots \Lb_{-\kb_{\bar{m}}}\,.
}[]
In particular $L_{-2}\unit = T$ and $\Lb_{-2}\unit = \Tb$ are the left and right components of the stress tensor. In order to compute the coefficients $a_\CO$ we will first need the OPE coefficients for $\phi\times\phi$. Following the algorithm explained in \cite{DiFrancesco:1997nk} we define
\eqn{
f_j = \sum_{|Y|=j}\beta_Y\,L_{-Y}\psi\,,
}[]
where $|Y|= k_1 + \cdots k_m$ and $\psi$ is assumed to be the lowest state of a non-degenerate Verma module with weight $h$. Then the covariance equations
\eqn{
L_1\lsp f_j - (h+j-1) f_{j-1} = 0\,,\qquad L_2 \lsp f_j - (h_\phi+h+j-2) f_{j-2} =0\,,
}[]
can be solved inductively to obtain the $\beta_Y$ to an arbitrary level. The determinant of this system is the Kac determinant, so the solution is always unique for non a degenerate module. In our case of interest $\psi=\unit$ is degenerate, so we need to set to zero its null state first: $\beta_1 = 0$.
\par 
The one-point functions are definitely more involved. We need to compute the transformation law for an arbitrary field---not necessarily a Virasoro primary---under the exponential mapping $z(w) = \exp(2\pi w)$. The general expression will involve sums of products of derivatives of the Schwartzian derivative
\eqn{
(Dz)(w) = \frac{z'''(w)}{z'(w)} - \frac32 \left(\frac{z''(w)}{z'(w)}\right)^2\,.
}[]
From the results of \cite{Gaberdiel:1994fs} we can write the transformation of a Virasoro descendant $L_{-Y}\psi$ as
\eqn{
z'(w)^{L_0} \prod_{n=1}^\infty e^{T_n(w) L_n}\,L_{-Y}\psi = v_Y(z^{(i)}(w))\,\psi + (\mbox{descendants of $\psi$})\,,
}[eq:transformationlaw]
where the $T_n(w)$ are defined recursively in \cite{Gaberdiel:1994fs}. In particular $3!\lsp T_2(w)$ is $(Dz)(w)$. Taking the expectation value of \eqref{eq:transformationlaw} with $\psi = \unit$ results in all descendants on the right hand side vanishing and thus
\eqn{
\vev{L_{-Y}\unit}_\beta = v_Y(z^{(i)}(w))\,.
}
We explicitly computed $v_Y$ and $\beta_Y$ up to level $|Y|=10$ for an arbitrary central charge and checked that the coefficients $a_\CO$ obtained by Taylor expanding \eqref{eq:twopftwod} agree with this formula
\eqn{
a_{\CO_{h,\hb}} = \sum_{|Y| = h}\sum_{|\Yb|= \hb} \beta_Y\lsp\beta_\Yb\lsp v_Y\lsp v_\Yb\,.
}[]

\Bibliography[./biblio]
\end{document}